\documentclass[journal=nalefd,manuscript=letter,layout=onecolumn,email=true]{achemso}
\usepackage{lineno,hyperref}
\usepackage{comment}
\usepackage{lscape}
\usepackage{xcolor}
\modulolinenumbers[1]

\title{Effects of surface chemical modifications on the adhesion of metallic interfaces. An high-throughput analysis.}

\author{E. Poli}
\affiliation{Department of Physics and Astronomy, University of Bologna, 40127 Bologna, Italy}
\email{emiliano.poli2@unibo.it}
\author{M. Cutini}
\affiliation{Department of Physics and Astronomy, University of Bologna, 40127 Bologna, Italy}
\author{M. A. M. A. Nosir}
\affiliation{Department of Physics and Astronomy, University of Bologna, 40127 Bologna, Italy}
\author{O. Chehaimi}
\affiliation{Department of Physics and Astronomy, University of Bologna, 40127 Bologna, Italy}
\author{M. C. Righi*}
\affiliation{Department of Physics and Astronomy, University of Bologna, 40127 Bologna, Italy}
\email{clelia.righi@unibo.it}

\begin{document}
\nolinenumbers

\begin{abstract}
\nolinenumbers
Chemical interactions between two surfaces in contact play a crucial role in determining the mechanical and tribological behavior of solid interfaces. These interactions can be quantified via the adhesion energy, that is a measure of the strength by which two surfaces bind together.
Several works in literature report how the presence of chemisorbed atoms at homo- and heterogeneous solid-solid interfaces drastically change their proprieties. A precise evaluation of how  different species at solid contacts modulates their adhesion would be extremely beneficial for a range of different technological fields: from metallurgy to nuclear fusion.
In this work we have used and high-throughput approach to systematically explore the effects of the presence of non-metallic elements, at different concentrations, on the adsorption and adhesion energies of different homogeneous metallic interfaces. 
Together with the databases for the adsorption and the adhesion energies, we calculated several other properties such as the charge transferred at the interface, the d-band edge shift for the substrate the Bond order and the interfacial density redistribution for the hundreds of systems analyzed. 
These values were used to define different trends with respect to chemical and concentration parameters that could be useful for the development of engineered interfaces with selected properties. In particular we noticed how the substrate with low filling of d-band are the most prone to adsorb ad-atoms and how the adsorption of almost all non-metallic elements decreases the adhesion energy of solid interfaces, particularly in the case of Fluorine. Carbon and Boron were the only two ad-atoms species that showed an opposite trend increasing the adhesion energy instead.
\end{abstract}

\section{Introduction}
    Solid-Solid interfaces are ubiquitous in natural and technological processes\cite{intro1,intro2,intro3}. The mechanical behaviour of such interfaces is dictated by an interplay between the macro\cite{interplay-macro} and microscopic properties\cite{interplay-micro} of the surfaces comprising the contact. In fact, while the physical processes that govern friction are not fully understood \cite{frict1} it is known that the origins of frictional interactions are the result of the shearing of microscopic junctions due to nano-asperities present at contacting surfaces\cite{Medina2014,Muser2017}. In this context adhesive interactions are extremely relevant given the high surface to volume ratio at nano-asperity contacts. Adhesion is defined by the atomistic interactions of the mated surfaces of which some relevant aspects are chemical composition, crystal orientation and stacking, presence of adsorbates, impurities, defects and segregates \cite{siegel_PRB2002, Feldbauer_2018, tripathi_2018,cornil_2020, wang_2021}. 
    In particular the presence of atomic adsorbates at metallic interfaces can modify the tribological and mechanical properties of interfaces with relevant implications in many technological fields such as metallurgy\cite{rabinowicz_1971}, nuclear fusion generator\cite{Matejicek-2013,Heuer-2019,Heuer-2019bis}, automotive, catalysis\cite{adh_catal}, clean fuel deployment\cite{Hyd_emb}. The phenomena influenced by the presence of adsorbates ranges from hydrogen embrittlement of steels\cite{Hyd_emb,Hyd_emb2} to superlubricity\cite{superlub} to electrode adhesion\cite{Elec_adh} to different substrates and many more.
    A deeper knowledge of the effects of different adsorbates should allow the prediction and optimization of the mechanical and the tribological behaviour of metallic interfaces, opening the possibility for applications where selected levels of adhesion and friction required is obtained through chemical engineering. The interplay between theoretical analysis and experimental evidence is pivotal
    in this direction.
    From a theoretical standpoint the advent of high-throughput software has dramatically enhanced the input that can be given to the experimental community to speed up/aide the discovery and testing of novel compounds with selected properties\cite{Curtarolo-2013,Haastrup-2018,Li_npj_comp_mat2022,Rosen_npj_comp_mat_2022,Hebnes_npj_comp_mat_2022}.  This approach hence has become a valuable tool for material advancements.
    In this paper we present an high-throughput study, based on an extension of the TribChem\cite{tribchem} program previously developed by our group, on how the absorption of non-metal atomic species at homogeneous metal-metal interfaces defines their adhesive and chemical properties. In particular we focus on how the presence of difference adsorbates influences the Adsorption and Adhesion energies given their relevance in defining the tribological properties of an interface\cite{PRL_rho}. We also analyse how these central quantities are correlated to several other chemical descriptors/properties.
    The study of these fundamental systems gives us a perspective of the fundamental interactions at plays at the boundary between metallic interfaces and is a first step in order to move towards the comprehension of the physics of more heterogeneus interfaces. 
    
\section{Computational Methods}

\subsection{Tribchem}
    The high-throughput screening of the influence on adhesion of adatoms at homogeneous metal interfaces has been carried out using the TribChem software developed by our group. The technical details of the code are outlined in a separate paper\cite{tribchem}. TribChem role is to automate the creation of input files and geometries, submit and run the theoretical calculations, and finally collect and analyze the generated data in external databases. 
    The code is mainly written in python and leverages Fireworks\cite{fireworks} and Atomate\cite{atomate} as workflow managers. In addition TribChem is interfaced with Pymatgen\cite{pymatgen} and MPInterfaces\cite{mpinterfaces} that are used to create, manipulate and process the Input/Output operations.
    The DFT calculations are run by Tribchem using the Vienna Ab initio Simulation Package (VASP\cite{VASP_one}). We used for all the simultions plane wave projector augmented wave (PAW\cite{VASP_due}) basis and the  Perdew-Burke-Ernzerhof\cite{PBE-1996} (PBE) -generalized gradient approximation of exchange correlation.
\subsection{Workflow outline}


    \begin{itemize}
        \item Tribchem retrieves the structure for the material of interest from the Material Project database.
        \item The bulk structures are optimised converging their lattice parameters via a fit with the Birch-Murnaghan equation of state.
        \item The DFT energy convergence w.r.t the k-point mesh and the plane-wave cutoff is then evaluated via an automated set of calculations.
        \item The slabs of the materials selected are the obtained cutting the bulk along a input/user specified direction. In our specific case the slabs were cut along the most stable orientation  of each crystalline structure: (111) for FCC crystal, (110) for BCC crystal and (0001) for HPC ones.
        \item The optimal slab thickness is selected via convergence of the surface energy.
        \item The adatom is added to the slab in the different high symmetry sites depending on the orientation of the surface (as an example: Long-bridge, short-bridge, or top for the (110) BCC crystals. The structures obtained are then relaxed until the convergence of the forces is minimized below 0.01 eV/\AA.
        \item The lowest energy configuration is then selected and the adsorption energy with respect to the clean slab is calculated and saved in the Adsorption Energies (E$_{ADS}$) database.
        \item The interface is then built reflecting the slab (and not the adatom) through a mirror plane parallel to the slab surface. The mirrored slab is then placed at a distance along the z-axis equal to the interlayer spacing of the bulk. The total supercell is finally formed adding 15 \AA\ of vacuum to decouple the Periodic Boundary Conditions (PBC) images along the z-direction. The structure is then relaxed along the z-axis until the convergence of the forces is minimized below 0.01 eV/\AA.
        \item The mirrored slab is then shifted laterally to match all the high symmetry positions of the lower slab and. Exploring the different shifts for which the high-symmetry points of the two surface match, and collecting the corresponding total energies, a 2-dimensional potential energy surface (PES) is built. The structure corresponding to the minima of the PES is then selected. Finally, this structure is fully relaxed until the convergence of the forces is minimized below 0.01 eV/\AA. 
\end{itemize}

\noindent
    The Adhesion energy (E$_{adh}$), charge density redistribution ($\rho_{dist}$), Bader charges \cite{Bader}, Density Derived Eelectrostatic Charges (DDEC) \cite{DDEC6}, Bond Order and d-band edge \cite{d-band} are then calculated and added to the respective databases. In particular in order to calculate the DDEC charges, Bond Order and d-band edges, we used the codes VASPKIT\cite{VASPKIT}, and CHARGEMOL\cite{DDEC6}. The capabilities of running post-processing analysis has not been fully implemented in the workflow yet and a more technical paper about these capabilities will follow.
    This full procedure is run for 1x1, 2x1, 2x2 cell dimensions (along the a and b cell axis respectively) considering in each case only one adatom per system and simulating in this way three different coverages, $\theta=$ 1.0 monolayer (ML), $\theta=$ 0.5 and $\theta=$ 0.25, respectively.

\subsection{Subtrates and adatoms choices}
    The substrates choice was driven by our previous work (and hence the availability of the E$_{adh}$ and $\rho_{dist}$ data) on the clean homogeneous interfaces of these metals. The series of substrates analysed comprises of: Ag, Al, Au, Cr, Cu, Fe, Ir, Mg, Mo, Ni, Pt, Ti, W, and Zn. These metals were chosen given their technological relevance and the fact that they constitute a representative set of transition metals with different d-orbital occupations. In addition Mg and Al were also considered as limit cases of metal which reactivity is dictated by s and p-orbitals.
    The adatom selection comprises of all reactive non-metals from the first second and third period (Cl excluded). Boron was also included given its relevance in previous tribological studies. 


\section{Results and Discussion}

 The database of the adsorption energies for the different substrates with the different adatoms adsorbed on top and the one of the adhesion energies of the interfaces with the adatoms intercalated are shown in Figure\ref{fig:silica-models1} A and B respectively. The data are reported for all the three different coverages considered. A more detailed picture of each individual system (i.e. each adatom on every substrate) can be found in Fig S1 for the adsorption energies and Fig S2 for the adhesion energies.
The quantities used to compile these databases are defined as follows:

\begin{equation}
    \label{eqn:EADS}
    \ \ E_{ADS}=\frac{(E_{SLAB+nAd}-nE_{Ad}-E_{SLAB})}{n} 
\end{equation}

Where E$_{SLAB+nAd}$ is the total energy of the substrate covered by n adatoms, E$_{Ad}$ is the energy of the isolated atom and E$_{SLAB}$ is the energy of the clean substrate).

\begin{equation}
    \label{eqn:EADH} 
    \ \ E_{ADH}=\frac{1}{A}(E_{12}-E^{1}_{Ad}-E^{2}) 
\end{equation}

Where A is the supercell area, E$_{12}$ si the total energy of the two slabs and the adatom in contact, E$^{1}_{Ad}$ is the energy of the isolated lower slab with adsorbed the adatom and E$^{2}$ is the energy of the isolated upper slab.

\begin{equation}
    \label{eqn:RHO} 
    \ \ \rho_{dist}=\frac{1}{2z_{0}}\int_{-z0}^{z0} \left| \frac{\rho_{12}-\rho_{1}^{Ad}-\rho_{2}}{A} \right| dz
\end{equation}

Where 2z$_0$ is the interface distance, $\rho_{12}$ is the planar average of the interface charge density $\rho_{1}^{Ad}$ 
is the planar average of the charge density of the bottom slab with the adsorbed adatom and $\rho_{2}$ is the planar average of the charge density of the bottom slab for each specific system. $\rho_{dist}$ is a metric to quantify the accumulation of charge at the interface \cite{PRL_rho}.

\begin{equation}
    \label{eqn:d-edge} 
    \ \ \epsilon_d=\frac{\int_{-\infty}^{\infty}{n_d (\epsilon})\epsilon d\epsilon}{\int_{-\infty}^{\infty}{n_d (\epsilon}) d\epsilon}
\end{equation}

Where $\epsilon_d$ is the center of d-projected DOS $\epsilon$ represents the Kohn–Sham eigenvalues and n($\epsilon$) is the calculated total d-PDOS.\\
We also calculated for each system the bader\cite{Bader} and DDEC charges and the DDEC bond order. 
DDEC is an atom in molecules charge partitioning where the total electron density (n(r)) is partitioned into overlapping atomic densities (n$_i$(r)). 

\begin{equation}
    \label{eqn:ddec} 
    \ \ n_i(r)=\frac{w_i(r)}{\sum_{k}(w_k(r))}n(r)
\end{equation}

The atomic partial charges are then computed by integrating the atomic electron densities over all space

\begin{equation}
    \label{eqn:ddec2} 
    \ \ q_i=z_i-N_i=Z_i-\sum{n_i(r)d^3r}
\end{equation}

where N$_i$ is the number of electrons assigned to atom i and z$_i$ is its effective nuclear charge. Using the same approach, higher-order atomic multipoles may be computed as first-order, second-order, n$^{th}$-order moments of the atomic electron densities.The weighting function in the DDEC method is described so that the atomic weights are simultaneously optimized to resemble the spherical average of n$_i$(r) and the density of a reference ion of the same element with the same atomic population N$_i$ . Following  this approach, the assigned atomic densities rapidly reach a converging multipole expansion of the QM electrostatic potential and the resulting atomic populations are chemically reasonable. 
Building on this charge partitioning the Bond Order of the different atoms can be calculated.

A more detailed explanation of these method is beyond the scope of this paper an the interested reader is remanded to Ref.\cite{DDEC6,DDECBO}

\subsection{Adsorption Energy Analysis}
\begin{figure}[H]
    \centering
    \includegraphics[width=1.0\textwidth]{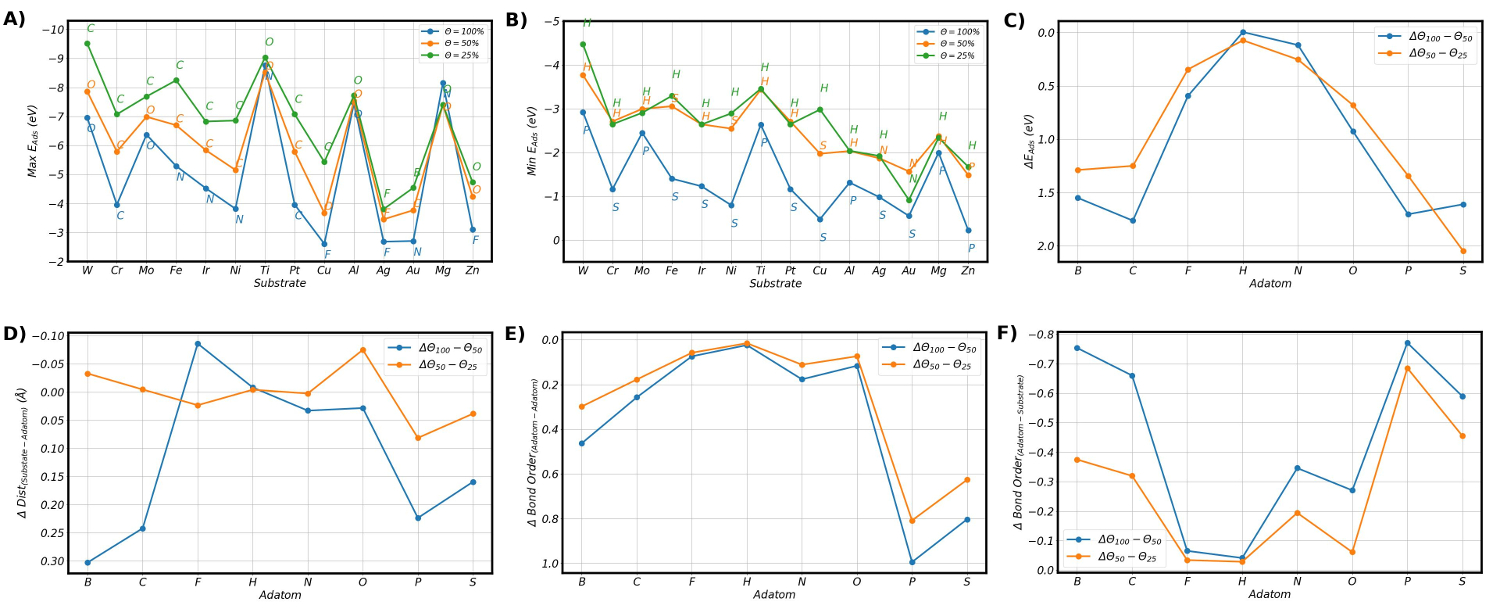}    
    \caption{Panel A: Maximum Adsorption Energy (E$_Abs$) for each substrate considered with respect to the coverage. The adatom for which the Maximum Adsorption Energy is obtained is specified in the label above the data-point. Panel B: Minimum Adsorption Energy (E$_Abs$) for each substrate considered with respect to the coverage. The adatom for which the Maximum Adsorption Energy is obtained is specified in the label above the data-point. Panel C: Average difference in Adsorption Energy between different coverage for all the adatoms considered. Panel D: Average change in distance from the substrate between different coverage for all the adatoms considered. Panel E: Average change in Bond Order between adatoms considering different coverage. Panel F: Average change in Bond Order between adatoms and the substrate considering different coverage.}
    \label{fig:silica-models2}
\end{figure}

Several general trends are observable regarding the adsorption energies:\\
i) The substrates with low filling of d-band are the most prone to adsorb adatoms (see Figure\ref{fig:silica-models1}, top panel). This trend is consistent with d-band centre theory for adsorption on metals for which the higher the d states are in energy relative to the Fermi level, the higher in energy the antibonding states are and the stronger the bond. This explain the lower propensity of noble metal and Zinc to adsorb different species \cite{Hammer-1995a,Hammer-1995b} with energies in the order of ~4 - 2 eV for the 25\% coverage, ~3 - 2 eV for 50\% coverage and ~2 - 1 eV for 100\% coverage. Conversely, the adsorption energies for the adatoms on the other substrate increase with the decrement of the d electrons. These trends were all confirmed by the d-band edge values calculated for all system (Table 1 Supporting information). 
Aluminum (that has only one p electron) shows adsorption energies that are comparable in magnitude with the ones of low filled d-bands substrates. It also shows (and Magnesium too) similar energetic trends to the low filled d-bands substrates with a high propensity to adsorb C, O and N. The case of Magnesium is difficult to evaluate since for several systems the adsorbed species percolates into the surface structure, in particular C, N and O. This is probably due to the high reactivity of the alkaline-earth metals and the large lattice parameter along z of Mg that allows for easier percolation. \\
ii) The most favorable adatom to adsorb depends both by the substrate and the coverage of the system(see Figure \ref{fig:silica-models2}a). The only three systems for which the coverage does not influence the most favorable species adsorbed are: Pt/C, Ag/F and  Al/O. In general the species with highest Adsorption Energy are the ones with smaller atomic radii exception made for hydrogen (i.e. C,N ,O and F). It is notable how at higher coverages the adsorption of Fluorine becomes more advantageous with respect the other adatoms.\\
iii) H is the least energetically favourite species to adsorb on the different substrates except for high concentrations where S and P are the least favourite especially on coniage metals. Figure \ref{fig:silica-models2}b in fact highlights how at $\Theta=0.5$ and $0.25$ H is generally the least favoured adatom on the different surfaces.\\
iv) Figure \ref{fig:silica-models2}b show the average change in Adsorption energy passing from $\Theta=1.0$ to $0.5$ and from  $\Theta=0.5$ to $0.25$. It can be seen that for atoms with smaller atomic radii the changes in E$_{ads}$ are smaller (i.e. the gain in energy due to adsorption is less coverage dependent). In particular for Hydrogen the adsorption energies remain on average the same, in the concentration range analysed. Boron, Carbon, Sulfur and Phosphorous E$_{ads}$ show instead a marked dependence by the coverage. Decreasing the coverage in these cases in fact leads to a drastic increase of the adsorption energies that seems to be correlated to a widening of the distance between the ad-atoms from the surface (see Figure \ref{fig:silica-models1}, bottom panel). Figure \ref{fig:silica-models2}d shows in particular how the B, C, P and S are on average ~0.3-0.2 \AA\ further away from the substrate at $\Theta=1.0$. For the other species the distance from the substrate remains almost the same exception made for F. These trends can in part be explained calculating the Bond Order between the adatoms and the different species present in the system. (see Figure \ref{fig:silica-models2}e,f)  For B, P, S (and in a more limited way C) the increment in coverage leads to an increment in the Bond Order of the adatoms with their periodic images (Figure \ref{fig:silica-models2}e) and a decrement of the total Bond Order with the atoms comprising the substrate (Figure \ref{fig:silica-models2}f). This implies that when the concentration of the substrate at the interface is increased the species with larger atomic radii tend to form stabilising bonds between their adatoms and decrease the interaction (form less bonds) with the substrate. To further corroborate this observation we calculated the Pearson correlation coefficients between the total Bond Order considering only adatoms-adatoms bonds and the different adatom atomic radii (i.e. B,C,F,H,N,O,P,S). The Paerson correlation coefficients are positive and above 0.2 for B(0.298), C(0.203), P(0.808) and S(0.625), indicating a strong positive correlation between these quantities, hence confirming our observation. For N the value is 0.115 indicating a somehow weaker correlation, while for H(0.015),F(0.038) and O(0.043) the coefficients are all lower that 0.05 indicating almost no correlation between the strength of interactions between adatoms (hence their coverage) and their Atomic radii. \\
v) Regarding the Bader and DDEC charge analysis (see Table 2 and 3 in the Supporting Information, showing the net charge transferred to the adatom) it can be observed that systems presenting higher adsorption energies also present bigger charge transfers between the substrates and the adatoms. To verify this trend for the whole database we calculated the Pearson correlation coefficients between E$_{Ads}$ and these two quantities (i.e. Bader and DDEC cahrges). We obtain a coefficient of 0.64 for the Bader charges and 0.70 for the DDEC charges, confirming the strong positive correlation between these quantities. The charge transfers seems to be especially significant in the two non-transition metal case and Titanium. It can be also observed, from the Bader charge analysis, that for Boron and Phosporous in some limited cases the electronic density is transferred from the adatom to the metal surface. This behaviour reflect the fact that these two species are the ones with the lowest electronegativity of the set we considered for the adatoms. This behaviour is observed also using the DDEC charge analysis but only for Phosphorous (although the Boron net charge values calculated are very near the zero). This discrepancy is probably due to the different charge partitioning of the two algorithms.

\subsection{Adhesion Energy Analysis}
\begin{figure}[H]
    \centering
    \includegraphics[width=0.95\textwidth]{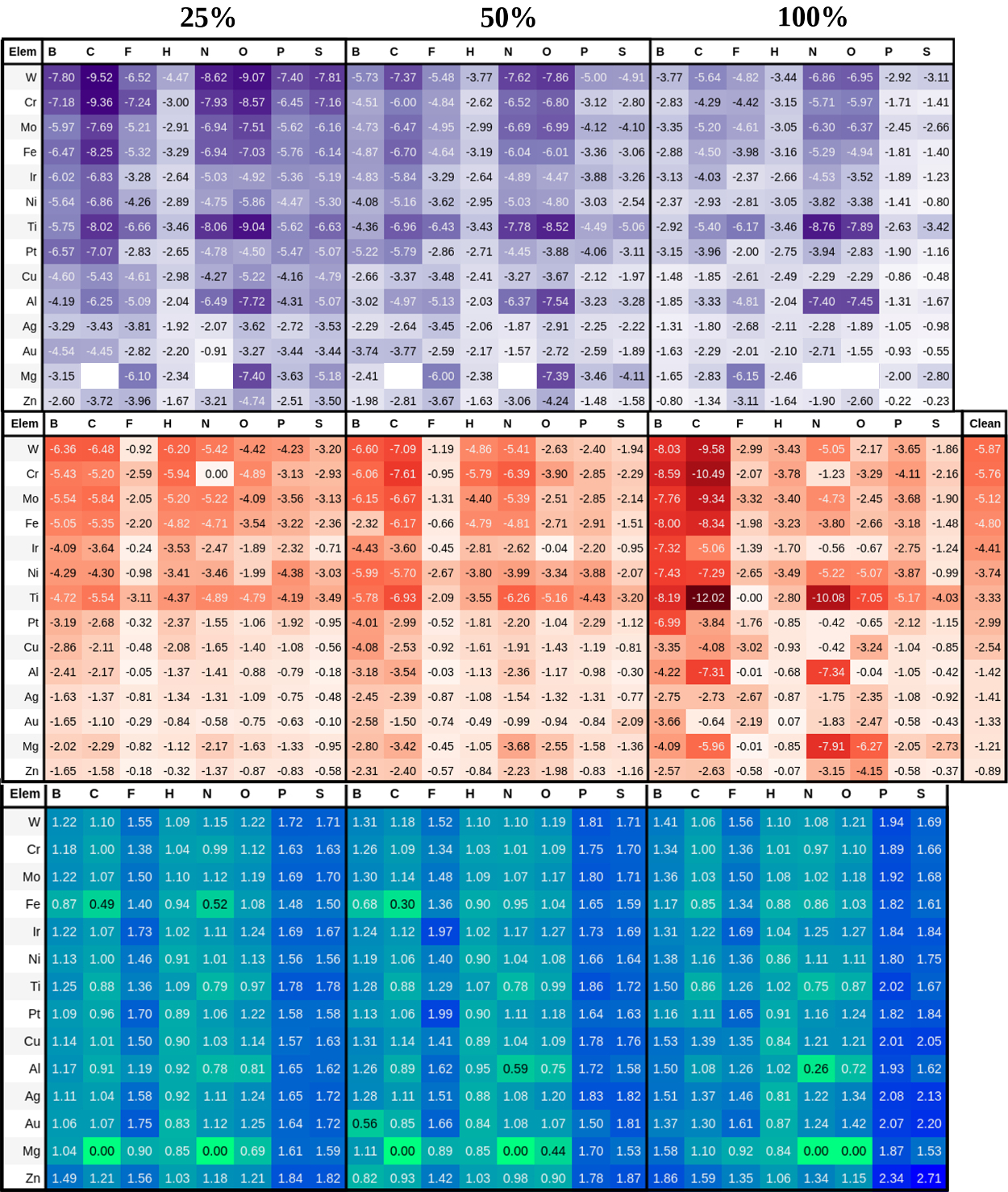}
    \caption{Adsorption energy in eV (Top panel), adhesion energy in (J/m$^{2}$) (Middle Panel) and vertical distance along z between the adatom and the xy-plane of the substrate slab in \AA (Bottom Panel) for 25\%, 50\% and 100\% coverages. Empty squares indicates when the atomic adsorption is not stable. In this cases, the adsorbate percolates into the surface structure. For the adhesion energies the clean interfaces values calculated in Ref. \cite{tribchem2} are reported in the Column "Clean".In both cases darker shades are related to higher adsorption and adhesion energies. for the adsorption systems at coverage $\Theta$=0.25,0.5,1.0. Squares with 0 values indicates when the atomic adsorption is not stable. In this cases, the adsorbate percolates into the surface structure.}
    \label{fig:silica-models1}
\end{figure}

\begin{figure}[H]
    \centering
    \includegraphics[width=1.0\textwidth]{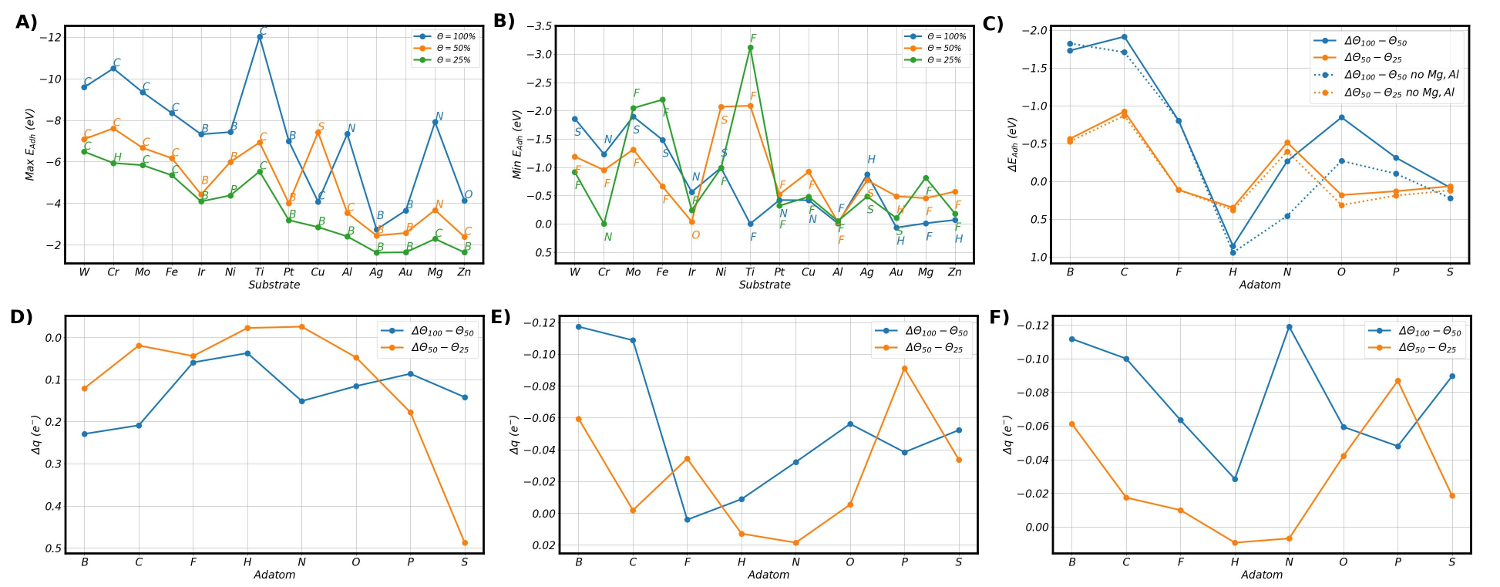}    
    \caption{Panel A: Maximum Adhesion Energy (E$_Adh$) for each substrate considered with respect to the coverage. The adatom for which the Maximum Adsorption Energy is obtained is specified in the label above the data-point. Panel B: Minimum Adhesion Energy (E$_Adh$) for each substrate considered with respect to the coverage. The adatom for which the Maximum Adsorption Energy is obtained is specified in the label above the data-point. Panel C: Average difference in Adhesion Energy between different coverage for all the adatoms considered. The dotted line represent the same analysis executed without considering the data obtained for Mg and Al (i.e. non transition metal). Panel D: Average change in the adatom charge (e$^-$) between different coverage for all the adatoms considered. Panel E: Average change in the down slab (i.e. the one with the adsorbed adatom) charge (e$^-$) between different coverage for all the adatoms considered. Panel F: Average change in the up slab charge (e$^-$) between different coverage for all the adatoms considered. All the charge analysis reported in this graphs are derived using Bader charges. The graphs obtained using the DDEC charges can be found in the Supporting Info.}
    \label{fig:silica-models3}
\end{figure}

Regarding the adhesion energy database we notice that:\\
i)The presence of adatoms leads in almost all cases to a reduction of the Adhesion energy (see Figures \ref{fig:silica-models1}B ). The important exceptions to this trend are Boron and Carbon that at low $\Theta$ slightly increase the Adhesion Energy and at higher coverage highly increase it. From Figure\ref{fig:silica-models3}a it can be seen that for almost all cases (Mg, Zn and Al excluded) at high concentrations Boron and Carbon are in fact the adatoms that result in the highest adhesion energy. Both these species are known for their flexibility in forming bonds \cite{BoronB1,BoronB2,CarbonB,CarbonB2}, hence we can hypothesize that at a certain concentrations such behavior could lead to the formation of stronger interactions at the interface w.r.t the other adatoms. However, given the recent literature highlighting the lubricant properties of graphene \cite{Graphlubr,Graphlubr2} is important to specify how the systems described in our simulations present a very different chemistry from the graphene case. In fact in our system the carbon atoms are adsorbed at distances where they cannot form stable bonds hence they are in a less stable configuration and more prone to interact with the substrate and transfer charge/electronic density. Our systems geometries (at least using the current workflow setup) are more suited to be compared to the addition of additive in alloys and metalworking. Several studies can be found where the addition of Boron and Carbon lead to an increment of the material strength, toughness, and high wear resistance\cite{BCFeAdh1,BFeAdh1,BoronAdh1,BoronAdh2,CNAdh,CTiAdh1}. Many of these works focused on Iron based systems\cite{BCFeAdh1,BFeAdh1} but even the other examples show that as long as the presence of these elements is interstitial all the previously listed properties are enhanced. In particular boron is nowadays commonly added to many structural alloys primarily as a grain-boundary strengthener\cite{BoronAdh1,BoronAdh2}.\\
ii) Figure\ref{fig:silica-models3}b shows instead that the lowest Adhesion energies are obtained for systems with Fluorine or Sulphur adsorbed at the interface. This trend reflect the fact that sulfur\cite{sulfurlub,sulfurlub2} and fluorine compounds are at the basis of several industrial lubricant additives. Several reports in literature also suggest F as promising adhesion-reducing element at interfaces \cite{fluorinelub2,flurinelub,fluorinelub3} and could be also connected to the hydrophobicity of fluorinated interfaces (such as Teflon). Our findings also confirm previous results for S that is know for its lubricating properties, especially at Fe interfaces \cite{FeS}.\\
iii) It is interesting to observe how the difference in coverage plays a less clear role for adhesion interactions  w.r.t. the adsorption energies case (see Figure\ref{fig:silica-models3}c). In fact for F, N, O and P there is on average an increment in adhesion energy going from $\Theta=0.5$ to $\Theta=1.0$ even if the overall E$_{Adh}$ value remains lower than in the clean interface case. On the contrary, the adhesion energy is lowered if we increase the coverage passing from $\Theta=0.25$ to $\Theta=0.5$ (exception made for B, C and N). The only two adatom species for which the Adhesion decrease constantly with and increment of  the coverage are Sulfur and Hydrogen. Such trend has relevant technological implications since hydrogen embitterment is a important challenge to resolve in order to enable hydrogen to become a wide-spread fuel option \cite{HEmbrit}. 

iv) A possible explanation/hint of the reasons behind these multivariate trends of the E$_{Adh}$ with respect to the coverage can be deduced by the charge analysis of the different components of the systems simulated. Figure \ref{fig:silica-models3}d-f show the average change in the net Bader charge of the adatom (Panel D), down slab (Panel E) and up slab (Panel F) w.r.t to the different change in surface decoration. Regarding the adatom net charge it can be seen that, excluded Hydrogen and Nitrogen with coverage passing from $\Theta=0.25$ to  $\Theta=0.5$, for all cases the decrement in coverage leads (on average) to a gain of negative charge on the adatom. Consequentially, this accumulation of electronic density on the adatoms generate a loss in negative charge on the up and down slabs (i.e. the slab on which the adatoms are adsorbed). However, the redistribution of the charge between the two slab is not symmetrical and leads to non trivial electrostatic pattern at the interface. A characteristic example is given by Fluorine where we always observe a loss in Adhesion energy w.r.t. to the clean interface. However in many of these systems passing from $\Theta=0.5$ to the monolayer leads to and increment in E$_{Adh}$ where going from $\Theta=0.25$ to $\Theta=0.5$ results in the opposite effect. From Figure \ref{fig:silica-models3}d-f it can be seen that the F adatom loses on average ~0.1 e$^-$ passing from coverage 0.25 to 0.5 and from 0.5 to 1. In response (on average) the down slab gains electron density going from $\Theta=0.25$ to $\Theta=0.5$ but not going from half coverage to the monolayer (~0 e$^-$ gained). The up slab gains electron density in both the increments in concentration but for the one passing from $\Theta=0.5$ to $\Theta=1$ the average negative net charged obtained is bigger (i.e. ~0.07 e$^-$). This results in a less negatively charged adatom monolayer/ less positively charged up-slab and hence in a diminished repulsive electrostatic interaction. This smaller interaction probably leads to an increment E$_{Adh}$. The F case is just an example to highlight how the adhesion energy is modulated case by case by the microscopic environment at the interface. This observation further underlines the necessity forcareful atomistic simulations of the interface environment/morphology.

\subsection{Interfacial density redistribution analysis}
\begin{figure}[H]
    \centering
    \includegraphics[width=1.0\textwidth]{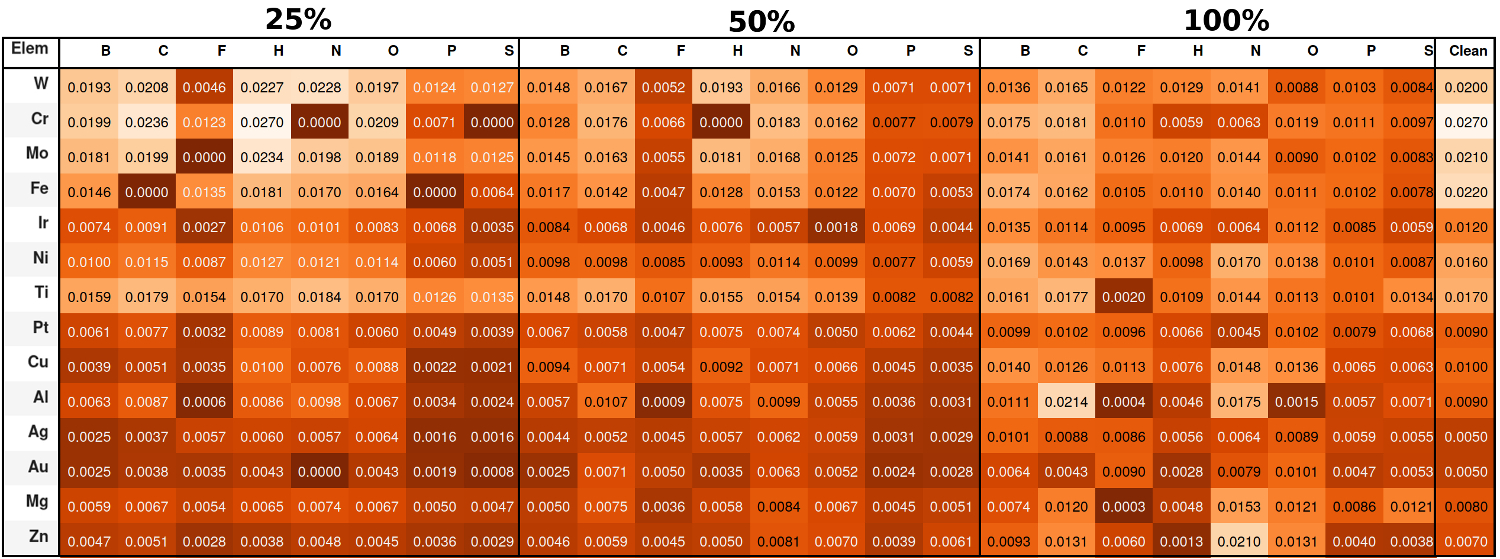}
    \caption{Charge redistribution factor $\rho_{redist}$ in $e^{-} Ang^{-3}$ for the adhesion systems at coverage $\Theta$=0.25,0.5,1.0. The clean interfaces values calculated are reported in the Column "Clean".}
    \label{fig:silica-models4}
\end{figure}

The Interfacial density redistribution should aid us in further explore the electronic density redistribution at the interface. $\rho_{redist}$ shows similar trends (see Image \ref{fig:silica-models4}) to the adhesion energy one w.r.t to the Adatom species in fact, for all the adatoms where the adhesion energy was drastically reduced (P,S and F) we can observe how the $\rho_{redist}$ values are smaller. This indicates a smaller redistribution of the charge density at boundary that in previous work has been correlated to lower adhesion and hence friction \cite{PRL_rho}. $\rho_{redist}$ values generally remain similar at lower concentrations ($\Theta= 0.25 and 0.5$) and than increase at full coverage. This behaviour could be dictated by the fact that the higher the concentration of the adatoms the bigger (on average) is the charge transferred with the two slabs (see Figure\ref{fig:silica-models3}). The only two species that don't follow this trend are H, N and to a lesser extent O. In particular the incremental presence of hydrogen decreases the electronic redistribution between metallic slabs. Almost all the values of $\rho_{redist}$ calculated for the 25-50\% concentrations are lower with respect to their Clean reference. This in principle respect the correlation observed between E$_{Adh}$ and $\rho_{redist}$ in previous work \cite{PRL_rho,tribchem2}. In fact the lower the adhesion the lower the charge density redistribution. However for B and C at $\Theta = 0.5$ decoration the $\rho_{redist}$ value is smaller that the clean interface one but E$_{adh}$ is bigger. This deviation from the trend is even more evident at full concentration where especially for d-metal with low filled orbitals. These deviations support our previous observations for the Adhesion Energies and further highlights how the presence,  concentration geometry of the adatoms has to be factored in the analysis interfacial charge density redistribution. In fact it seems that the concentration of the adatoms together with the charge transfer they induce and the consequent electrostatic interactions at the nanoscale result in a non-trivial energetic picture with respect to the Adhesion interaction.

\section{Conclusions}
In summary, we used an high-throughput approach to the study of the effects of the presence of different species at several homogeneous metallic interfaces. This systematic approach allowed us to uncover/confirm several interesting trends:
\begin{itemize}
    \item The substrates with low filling of d-band are the most prone to adsorb ad-atoms. This trend is consistent with d-band centre theory for adsorption on metals.
    \item For P, S and B increasing the coverage leads to a drastic reduction of the adsorption energies that seems to be correlated to a widening of the distance between the ad-atoms from the surface and the formation of bonds between adatoms. 
    \item For H E$_{ads}$ varies much less with the change in coverage w.r.t all the other species analysed.
    \item Exception made for B and C the presence of non-metal adatoms always reduces the adhesion. Our adhesion results were confirmed/supported by several experimental results. Boron and Carbon are well known to be used in metal casting and alloying to improve material strength, toughness, and high wear resistance and Boron more specifically is used as a grain-boundary strengthener. Such correspondence between experimental literature and our results underlines the potential usefulness of our adhesion database to engineer material with modular toughness, wear resistant properties.
    \item P, S and in particular F are strong adhesion reducer. This trend was also supported by several works we found in literature, however especially for P and S the experimental studies were more focused on lubricant additives hence their effects on the sliding behaviour at the  interface should be further explored.
    \item Hydrogen present a unique behaviour respect to the other species probably due to its small atomic radius. Its increment in concentration at the interface in fact promote a clear reduction in adhesion. 
    \item The increment in coverage of the different adatoms lead to a non-trivial non-symmetric charge transfer picture at the interface that can give rise to different electrostatic regimes at the nanoscale.
    \item This non-trivial picture is also reflected in the $\rho_{redist}$ values. The relation between $\rho_{redist}$ and E$_{adh}$ observed in previous works seems to still be present. However the effects of the nature and concentration of the adatoms give rise to several exception that will need to be further analysed.
\end{itemize}
This work represents a first step in the methodical study of more realistic models of solid interfaces and their mechanical properties. Obviously the systems simulated here are still far from realistic conditions, temperature, pressure, heterogeneity, disorder and several other factors are missing. Nonetheless our workflow approach can be extended (with relative ease) to include some of these parameters and support further experimental applications. Nonetheless we found several experimental work upholding our findings. This aims to be the starting point to inform the experimental community about functional interfaces in adhesion applications. In fact our future goal is to further expand our workflow in order to consider a wider range of concentrations materials and heterogeneous contacts so that a well defined interface with pre-determined adhesion properties can be engineered/obtained staring from from ab-initio data instead that just empirical work.

\section{Supporting Information Available}
 The Supporting Information is available free of charge.
This contains the databases of the d-band edges calculated for the absorption systems, the database for the Bader charges of the Adatoms calculated for the absorption systems and the database for the DDEC charges of the Adatoms calculated for the absorption systems.

\section{Competing interests}
The authors declare no competing interests.

\section{Acknowledgements}
These results are part of the ”Advancing Solid Interface and Lubricants by First Principles Material Design (SLIDE)” project that has received funding from the European Research Council (ERC) under the European Union’s Horizon 2020 research and innovation program (Grant agreement No. 865633).

\bibliography{biblio-titlecase}%

\end{document}